\documentclass[12pt, a4paper]{article}

\usepackage{graphicx}

\usepackage{stmaryrd}
\usepackage{overcite}
\usepackage{verbatim}
\usepackage{achemso}
\usepackage{verbatim}

\usepackage{multirow}
\usepackage{amsmath}
\usepackage{amsfonts}
\usepackage{amsgen}
\usepackage{longtable}
\usepackage{lscape}
\usepackage{threeparttable}
\usepackage{supertabular}
\usepackage{multicol}
\usepackage{float}
\usepackage{multirow}
\usepackage{setspace}
\usepackage{rotating}
\usepackage{amssymb}
\usepackage{layout}
\usepackage{wrapfig}
\usepackage{subfigure}
\usepackage{subfig}
\usepackage{color, soul}

\begin{document}

\begin{center}

{\bf\large{Voltammetry at Porous Electrodes: A Theoretical Study}}

\hspace{2cm}

{\bf\large Edward~O.~Barnes$^a$, Xiaojun~Chen$^{a,b}$, Peilin~Li$^a$ and Richard~G.~Compton$^a$*}

*Corresponding author

$^a$Department~of~Chemistry, Physical~and~Theoretical~Chemistry~Laboratory, Oxford~University,
South~Parks~Road, Oxford, OX1~3QZ, United~Kingdom.
$^b$College~of~Sciences, Nanjing~Tech~University, Nanjing, 211816, P.~R.~China.

Fax:~+44~(0)~1865~275410; Tel:~+44~(0)~1865~275957.
Email:~richard.compton@chem.ox.ac.uk
\vspace{1cm}

\end{center}

NOTICE: this is the author’s version of a work that was accepted for publication in \emph{The Journal of Electroanalytical Chemistry}. Changes resulting from the publishing process, such as peer review, editing, corrections, structural formatting, and other quality control mechanisms may not be reflected in this document. Changes may have been made to this work since it was submitted for publication. A definitive version was subsequently published in \emph{The Journal of Electroanalytical Chemistry}, 720 (2014) 92-100 DOI: 10.1016/j.jelechem.2014.03.028.

\clearpage

\section*{Abstract}

Theory is presented to simulate both chronoamperometry and cyclic voltammetry at porous electrodes fabricated by means of electro-deposition around spherical templates. A theoretical method to extract heterogeneous rate constants for quasireversible and irreversible systems is proposed by the approximation of decoupling of the diffusion within the porous electrode and of bulk diffusion to the electrode surface.

\section*{Keywords}

Cyclic voltammetry; Diffusion within a sphere; Templated porous electrodes; Potential step chronoamperometry.

\clearpage

\section{Introduction}

The use of porous electrodes for electroanalytical measurements is growing because of the scope for altered voltammetric responses which may, under favourable circumstances, offer a more sensitive response and at a lower over-potential than conventional macro-electrodes where mass transport occurs via semi-infinite diffusion. Such electrodes are often made by using as a template layers of spheres, for example made of polystyrene, on the surface of a pre-existing macroelectrode. Metal, very often gold, is then electroplated around the spheres before the latter are removed, for example via dissolution in a suitable solvent. The created porous structure can then, in an idealised sense, be thought of as a three dimensional network of hollow metal spherical holes interconnected via tiny ducts which allow hole to hole transport and which are likely formed at the points of contact between the original templating spheres. Table \ref{POROUS USES} shows recent examples of the use of porous electrodes fabricated in this manner in electrochemical sensing, where both improved sensitivity and limits of detection are often claimed. These include the detection of species such as NADH\cite{Szamocki2006, Szamocki2007, Lenz2011}, glucose\cite{Chen2009, Chen2012, Chen2012a}, nitrobenzene\cite{Zhang2012} and hydrogen peroxide\cite{Wei2011}, in addition to a number of immunosensors\cite{Chen2008, Li2009b, Chen2010}.

The purpose of the present paper is to develop theory for porous electrodes constructed with the method described above and summarised in Fig \ref{SCHEME}. In particular we seek to identify the contributions to the voltammetric and chronoamperometric signals resulting from material diffusing to the top of the surface of the electrode from bulk solution and from material within the spherical holes of the porous structure. Qualitatively we anticipate that the latter will give rise, under suitable timescales, to a ``thin layer'' type response and that this can lead to a reduced overpotential. If a mixture of analytes in solution have similar formal potentials under bulk diffusion conditions alone, the additional thin layer element may be used to discriminate between them if they enter the porous layer to differing extents.\cite{Streeter2008c, Sims2010, Henstridge2010}.

In the following we employ a ``decoupling'' of the two modes of transport, diffusion to the electrode surface and transport within the porous layer, which has been shown experimentally to be the case for porous electrodes fabricated using drop casting of carbon nanotubes\cite{Punckt2013}.

\section{Theory}

Electrochemistry at a porous electrode is simulated. A simple, single electron transfer is considered:
\begin{equation}
\mathrm{A} \pm \mathrm{e^{-}} \rightleftharpoons \mathrm{B}
\end{equation}
The transfer is assumed to take place in solution with a large amount of inert supporting electrolyte to ensure that mass transport is diffusion only. 

\begin{figure}[ht]
\begin{center}
\includegraphics[width=\textwidth]{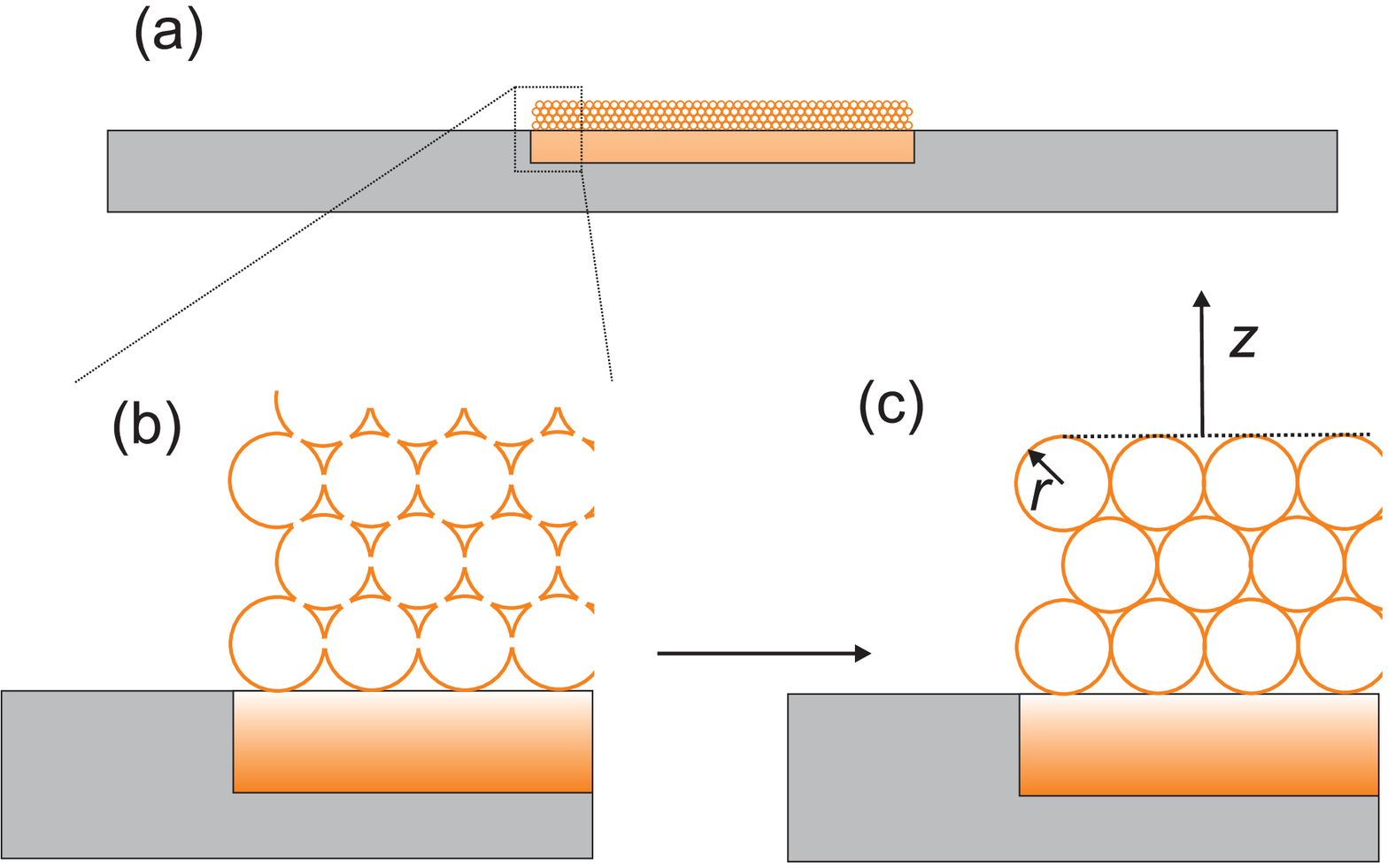}
\caption{Schematic cross section of a porous electrode consisting of a series of hollow spheres supported on a disc electrode (the size of sphere relative to the disc electrode is greatly exaggerated). (b) shows a zoomed in section, highlighting then interconnected nature of the spheres. (c) shows a depiction of the model used to simulate electrochemistry at the electrode.} \label{SCHEME}
\end{center}
\end{figure}

The porous electrode is considered to be a series of interconnected spherical voids within a conductive material. These electrodes may be fabricated by supporting a series of nanosized polystyrene spheres on a macroelectrode surface, and surrounding them with a layer of, for example, electro-deposited gold. The spheres are then dissolved away, leaving behind a network of interconnected holes, through which solution containing the analyte of interest may permeate. For simplicity, and to generate an idealised structure amenable to theoretical modeling, we assume the porous layer comprises hollow spheres formed from where the polystyrene were dissolved and tiny pores of negligible size linking adjacent spheres. This allows the entire porous layer to be flooded with electrolyte when immersed in a solution. This is shown schematically in Fig.\ref{SCHEME}, which shows a view of the spheres supported on an electrode (a), and a zoomed in view of the porous, interconnected hollow sphere structures (b). In this study, the electrode is modeled as a series of identical, hollow, independent conductive spheres (Fig. \ref{SCHEME} (c)). After the start of an experiment, it is assumed that negligible diffusion occurs either from sphere to sphere or between the spheres and the bulk. If the spheres are small enough relative the the radius of the disc they are supported on, then the top of the layer(s) of spheres can be considered a disc electrode. The reasons for this latter simplification are discussed next.

\begin{figure}[ht]
\begin{center}
\includegraphics[width=\textwidth]{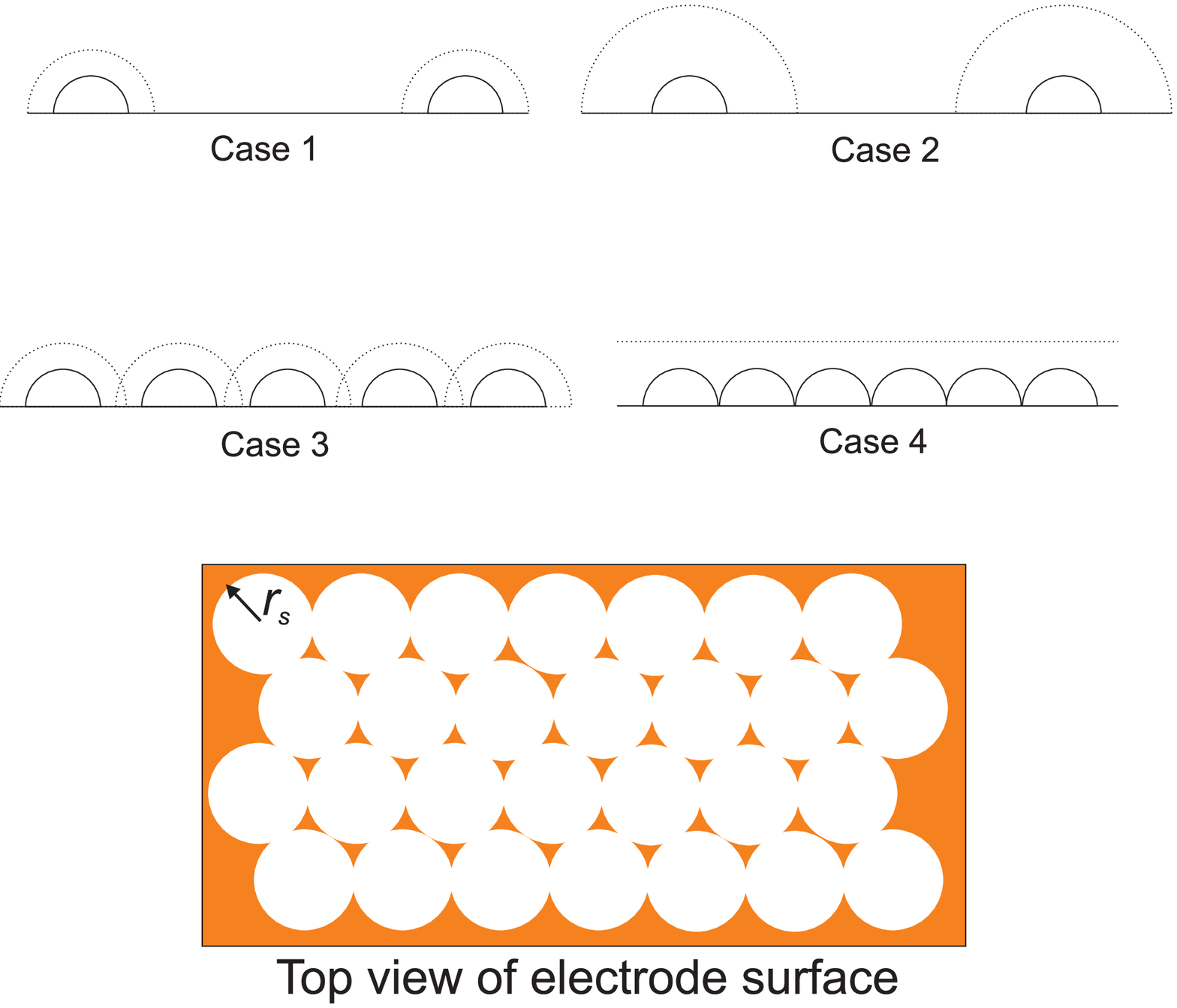}
\caption{Schematic diagram of the four cases for electrochemical responses at a micro- (or nano-) electrode array, and a top view of a porous electrode surface, showing the close packing of hemispherical hollows.} \label{CASES}
\end{center}
\end{figure}

The electrochemical behavior of flat micro- (and nano-) electrode arrays can be divided into four distinct cases\cite{Brookes2003, Davies2004, Davies2005b, Chevallier2005, Davies2003, Davies2005, Davies2005a}, as summarised in Fig. \ref{CASES}. In case 1, the timescale is short enough so that the diffusion layers are independent and do not overlap. This leads to linear diffusion and a Cottrellian response from the array as a whole. In case 2, the timescale is longer, and while the electrodes still have independent diffusion layers, these layers are now large compared to the electrodes, and convergent diffusion occurs. The response can be considered to be that of several independent micro electrodes. In case 3, the diffusion layers become comparable in size to the electrode separation, and begin to overlap. In case 4, the diffusion layers are very large compared to the electrodes and the distance between them. The diffusion layers merge into a single diffusion layer, and the array behaves as a single, much larger electrode, and Cottrellian (planar diffusion) behavior is again observed. By considering the top layer of spheres to behave as a case 4 array, it may be modeled as a macroelectrode. This will be valid provided at least several layers of spheres are used in the electrode construction and that $\sqrt{Dt}>r_s$, where $t$ is the timescale of the experiment, $D$ is the diffusion coefficient of the species interest and $r_s$ is the radius of the sphere. These conditions are thought to hold for the electrodes studied later in this paper.

Further to this assumption of modeling the top layer of spheres as a case 4 electrode array, we also assume that diffusion inside the hollow spheres is completely decoupled from the diffusion down to the top layer of spheres. We also assume that no diffusion occurs between individual spheres. This allows us to simulate the response inside a single sphere and and the response at a disc electrode. The single sphere result can be multiplied by the number of spheres present, and added to the disc response from the exposed top layer of spheres to give an overall response. This model is used to simulate two electrochemical experiments: potential step chronoamperometry and cyclic voltammetry, which are now discussed in turn.

\subsection{Simulation of chronoamperometry within a single sphere}

In a potential step chronoamperometric experiment, the potential applied to the electrode is stepped from a value where no electron transfer takes place, to a value where electron transfer occurs at a rate dictated by the mass transport of the chemical species to the electrode surface. The spherically symmetric environment inside a sphere can be reduced to a one dimensional system. The dependence of concentration of any species on time is given by Fick's second law in spherically symmetric space:
\begin{equation}
\frac{\partial{c_i}}{\partial{t}} = D_i\left(\frac{\partial^2{c_i}}{\partial{r^2}} + \frac{2}{r}\frac{\partial{c_i}}{\partial{r}}\right)
\end{equation}
where $c_i$ is the concentration and $D_i$ the diffusion coefficient of species $i$, and $r$ is the radial coordinate, equal to zero at the centre of the sphere, and equal to $r_e$ at the inside edge of the sphere. Symbols are defined in Table \ref{DIMENSIONAL}. Initial conditions are:
\begin{equation}
\text{$t = 0$, all $r$; } c_\mathrm{A} = c_\mathrm{A}^{*}
\end{equation}
Note that for single step chronoamperometry, as considered here, species B need not be considered.

At time $t=0$, the start of the experiment, the electrode potential is stepped, and the boundary condition at the electrode surface becomes:
\begin{equation}
\text{$t \geq 0$, $r=r_s$; }c_\mathrm{A} = 0
\end{equation}

At the centre of the sphere, a zero flux boundary condition is imposed on both species as a result of symmetry:
\begin{equation}
\text{$t \geq 0$, $r=0$; }\frac{\partial{c_\mathrm{A}}}{\partial{r}} = 0
\end{equation}

In order to simplify the model, dimensionless parameters are introduced. These are listed in Table \ref{DIMENSIONLESS}. Upon introduction of these parameters, the mass transport equation becomes:
\begin{equation}
\frac{\partial{C_i}}{\partial{\tau}} = D^{'}_i\left(\frac{\partial^2{C_i}}{\partial{R^2}} + \frac{2}{R}\frac{\partial{C_i}}{\partial{R}}\right)
\end{equation}
Dimensionless boundary conditions are listed in Table \ref{SPHERE BOUNDARY CONDITIONS}.

To calculate the current, a dimensionless flux at the sphere surface is calculated:
\begin{equation}
j_\mathrm{s} = \left(\frac{\partial{C_\mathrm{A}}}{\partial{R}}\right)_{R=1}
\end{equation}
This is transformed into a real, dimensional current, $I$ / Amps, by:
\begin{equation}
I_\mathrm{s} = 4\pi F r_s c_\mathrm{A}^{*}D_\mathrm{A}j_\mathrm{s}
\end{equation}

\subsection{Simulation of cyclic voltammetry within a single sphere}

The only change that need be made to the above model to instead simulate cyclic voltammetry inside a sphere is the electrode surface boundary condition, where Butler-Volmer kinetics are instead employed. Using Butler-Volmer kinetics\cite{Butler1924}, which can be shown to be a limiting form of assymetric Marcus-Hush theory\cite{Laborda2013}, the flux at the electrode surface is given by:
\begin{equation}
D_\mathrm{A}\left(\frac{\partial{c_\mathrm{A}}}{\partial{r}}\right)_{r=r_s} = k^0\left[c^0_\mathrm{A}\text{exp}\left(-\alpha\frac{F\left(E-E_f\right)}{RT}\right) - c^0_\mathrm{B}\text{exp}\left(\beta\frac{F\left(E-E_f\right)}{RT}\right)\right]
\end{equation}
where $\alpha$ and $\beta$ are transfer coefficients, and it is assumed that $\alpha + \beta = 1$.
Conservation of mass results in equal fluxes of species A and B:
\begin{equation}
D_\mathrm{B}\frac{\partial{c_\mathrm{B}}}{\partial{r}} = -D_\mathrm{A}\frac{\partial{c_\mathrm{A}}}{\partial{r}}
\end{equation}

Dimensionless boundary conditions are again summarised in Table \ref{SPHERE BOUNDARY CONDITIONS}. The dimensionless flux density and the real current are calculated the same as for chronoamperometry.

\subsection{Simulation of chronoamperometry and cyclic voltammetry at a disc electrode} \label{DISC CHRONO}

Chronoamperometry at a disc electrode is well characterised and may be calculated using the Shoup and Szabo equation \cite{Shoup1982, Paddon2007, Klymenko2004}. In our model of the porous electrode by hypothesis we decouple the current response from the interior of the spheres from that of the overall ``disc''. Thus there are two electrode radii (that of a sphere and that of the disc), which without due care will complicate the analysis of theoretical results in terms of dimensionless parameters. We therefore use the radius of the spheres as the basis for our set of dimensionless parameters. A dimensionless disc radius, $R_d$, must then be defined as:
\begin{equation}
R_d = \frac{r_d}{r_s}
\end{equation}
The Shoup-Szabo equation can then be expressed:
\begin{equation}
j_d = R_d\frac{2}{\pi}\left(0.7854 + 0.4432\left(\frac{\tau}{R_d^2}\right)^{-0.5} + 0.2145\text{ exp}\left[-0.3912\left(\frac{\tau}{R_d^2}\right)^{-0.5}\right]\right)
\end{equation}
and
\begin{equation}
I_d = 2\pi FD_\mathrm{A}c_\mathrm{A}^{*}r_sj_d
\end{equation}

Simulation methods for cyclic voltammetry at a macrodisc electrode have been well established. The relevant mass transport equation is:
\begin{equation}
\frac{\partial{C_\mathrm{i}}}{\partial{\tau}} = D^{'}_\mathrm{i}\left(\frac{\partial^2{C_\mathrm{i}}}{\partial{Z^2}}\right)
\end{equation}
Butler-Volmer kinetics are again used as the electrode surface boundary condition, now applied at $Z=0$ rather than $R=1$. The zero flux bulk solution boundary condition is applied at $Z_\mathrm{max}=6\sqrt{D^{'}_\mathrm{max}\tau_\mathrm{max}}$, where $D^{'}_\mathrm{max}$ is the largest (dimensionless) diffusion coefficient in the system and $\tau_\mathrm{max}$ is the total dimensionless time taken to run the experiment. This position has been shown to be well outside the diffusion layer\cite{Gavaghan1998, Gavaghan1998a, Gavaghan1998b}. Dimensionless boundary conditions, and the boundaries at which they apply, for a macrodisc are shown in Table \ref{DISC BOUNDARY CONDITIONS}.

The dimensionless flux at the disc, $j_\mathrm{d}$, (defined as):
\begin{equation}
j = \frac{I}{4\pi FD_\mathrm{A}c_\mathrm{A}^{*}r_s}
\end{equation}
is given by:
\begin{equation}
j_\mathrm{d} = \frac{R_d}{2}\left(\frac{\partial{C_\mathrm{A}}}{\partial{Z}}\right)_{Z=0}
\end{equation}

\subsection{Total current at a porous electrode}

We assume the electrochemical response at a porous electrode is the sum of the responses inside the hollow spheres (modeling the pores) and the disc like response from diffusion to the top layer of spheres. If there is negligible diffusion between individual spheres and the bulk during electrolysis, then the overall current will be the sum of the two responses:
\begin{equation}
I = I_s + I_d
\end{equation}
where $I_s$ and $I_d$ are the currents from the spheres and the ``disc'' respectively. Defining the total number of spheres as $N$, then in dimensionless terms:
\begin{equation}
j = Nj_s + j_d
\end{equation}
and
\begin{equation}
I = 4\pi FD_Ac_Ar_sj
\end{equation}

\subsection{Numerical Methods}

Numerical simulation of chronoamperometry and cyclic voltammetry necessitates the discretisation of the mass transport equations and boundary conditions in space and time, for which the Crank-Nicolson method is used\cite{Crank1947}. The equations are then solved simultaneously and implicitly over all space using the Thomas algorithm\cite{Press2007} to solve the large banded matrices produced.

Appropriate spatial and temporal grids must be defined to discretise and solve the equations over. For chronoamperometry, the temporal grid consists of a dense, regular mesh of points from $\tau=0$ up to some switching value, $\tau_s$. After this value, the temporal grid expands and becomes less dense. Mathematically:
\begin{eqnarray}
\tau_0 & = & 0\\
\tau \leq \tau_s \text{:} \quad \tau_k & = & \tau_{k-1} + \Delta_\tau \\
\tau > \tau_s \text{:} \quad \tau_k & = & \tau_{k-1} + \gamma_\tau\left(\tau_{k-1} - \tau_{k-2}\right)
\end{eqnarray}
For cyclic voltammetry, the temporal grid is defined in terms of the dimensionless potential. Each unit of $\theta$ is divided into $N_\theta$ evenly spaced points. The temporal grid then defined to be:
\begin{equation}
\tau_k = \tau_{k-1} + \frac{1}{N_\theta\sigma}
\end{equation}
For the spatial grid inside a sphere (for both chronoamperometry and cyclic voltammetry), the grid expands away from the electrode surface at $R=1$ after an initial step of $\Delta_R$. Mathematically:
\begin{eqnarray}
R_0 & = & 1\\
R_1 & = & 1 - \Delta_R\\
R_j & = & R_{j-1} - \gamma_R\left(R_{j-2} - R_{j-1}\right)
\end{eqnarray}
The final $R$ point is defined as zero, the centre of the sphere.

For a macrodisc, a similar grid is used, with an initial step size of $\Delta_Z$ from zero, and then expanding out to bulk solution:
\begin{eqnarray}
Z_0 & = & 0\\
Z_1 & = & \Delta_Z\\
Z_j & = & Z_{j-1} + \gamma_R\left(Z_{j-1} - Z_{j-2}\right)
\end{eqnarray}

Convergence studies found the following grid parameters sufficient to ensure simulated results were within 0.2\% of fully converged outcomes: $\Delta_\tau = 1\times 10^{-9}$, $\tau_s = 1\times 10^{-5}$, $\gamma_\tau = 1.0001$, $N_\theta = 100$, $\Delta_R = \Delta_Z = 1\times 10^{-5}$, $\gamma_R = \gamma_Z = 1.01$. Typical run times were of the order 5 to 10 seconds per simulation. The models were programmed in C++ and all simulations carried out on an Intel(R) Xenon(R) 2.26 GHz PC with 2.25 GB RAM.

\section{Simulated results and discussion}

\subsection{Chronoamperometry}

If we consider the chronoamperometric response inside the spherical holes of a porous electrode to be completely decoupled from that outside, then the total response will be a mixture of two kinds of diffusion; that occurring inside the spherical voids and that occurring at the ``disc like'' surface of the top layer of spheres shown schematically in Fig \ref{SCHEME}. Note that the number of layers of spheres supported on the disc electrode is arbitrary, subject to fulfilling the assumption identified above. The important parameter is the number of spheres, $N$.

\begin{figure}[ht]
\begin{center}
\includegraphics[width=\textwidth]{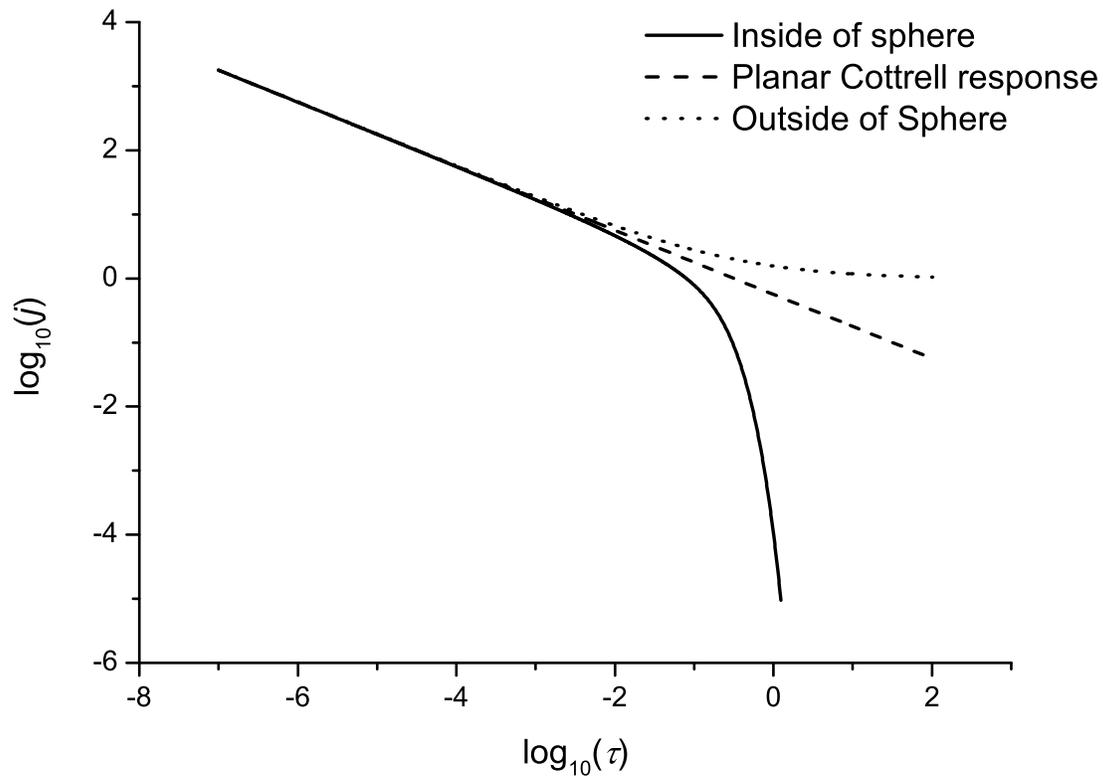}
\caption{Chronoamperometric responses (in dimensionless parameters) of the inside of a sphere (solid line), a macrodisc (dashed line) and the outside of a sphere (dotted line).} \label{CHRONO CASES}
\end{center}
\end{figure}

To understand these types of diffusion, chronoamperometric responses due to each kind are calculated and compared. Chronoamperometry is simulated inside a hollow sphere using the model outlined above, giving the dimensionless result shown by the solid line in Fig \ref{CHRONO CASES} (which shows dimensionless flux against dimensionless time in a log-log plot). This result is compared to known analytical results for an isolated planar macrodisc electrode, given by the Cottrell equation\cite{Cottrell1902}:
\begin{equation}
I_\mathrm{d} = \frac{nFAc\sqrt{D}}{\sqrt{\pi t}}
\end{equation}
which in the dimensionless form becomes:
\begin{equation}
j_\mathrm{d} = \frac{1}{\sqrt{\pi\tau}}
\end{equation}
This is shown as a dashed line in Fig. \ref{CHRONO CASES}. For comparison, the analytical equation for the chronoamperometric response \emph{outside} an isolated sphere is given by\cite{Compton2010}:
\begin{equation}
I_\mathrm{s'} = nFAc\left(\frac{\sqrt{D}}{\sqrt{\pi t}} + \frac{1}{r_e}\right)
\end{equation}
(where s' denotes the response outside of a sphere), given in dimensionless parameters by:
\begin{equation}
j_\mathrm{s'} = \frac{1}{\sqrt{\pi\tau}} + 1
\end{equation}
This is shown as a dotted line in Fig. \ref{CHRONO CASES}. At small values of $\tau$, all three cases show linearity in the log-log plot, corresponding to planar diffusion very close to the surface. In this limit, the diffusion is blind to the large scale shape of the electrode. As $\tau$ increases, however, differences become apparent. The response inside the sphere starts to drop off rapidly, as all the material inside the sphere is used up. This is in stark contrast to the well known response outside a sphere, which reaches a limiting steady state current due to efficient spherical diffusion. The ``infinite'' Cottrellian macrodisc response continues to be linear, and decreases steadily to zero.

By modeling a porous electrode as being the sum of a large number of hollow spheres and the overall disc, we can approximate the chronoamperometric response of the porous electrode. The contribution of the spheres is obtained \emph{via} simulation. The exposed surface of the top layer of spheres can be considered as a disc in which the conductive gold deposited around the upper most layer of spheres is sufficient to generate case 4 like behavior so that the porous electrode surface behaves like macrodisc. The basis for this is provided by extensive previous simulations which show that only a tiny amount of active surface is required to achieve diffusion control to the entire geometric disc surface\cite{Brookes2003, Davies2004, Davies2005b, Chevallier2005, Davies2003, Davies2005, Davies2005a}. Since
\begin{equation}
d \ll r_s \ll r_d
\end{equation}
(where $d$ is the separation between spheres) the diffusion layers around each top layer pore will overlap to a large extent and form a single, large diffusion layer. The contribution of the disc can be therefore obtained \emph{via} the Shoup and Szabo equation, as outlined in Section \ref{DISC CHRONO}.
\begin{equation}
j_\mathrm{d} = R_d\frac{2}{\pi}\left(0.7854 + 0.4432\left(\frac{\tau}{R_d^2}\right)^{-0.5} + 0.2146\text{ exp}\left[-0.3912\left(\frac{\tau}{R_d^2}\right)^{-0.5}\right]\right) \label{SHOUP AND SZABO}
\end{equation}
At any given time, the total dimensionless flux density will then by the sum of the flux from the (micro)disc given by Equation \ref{SHOUP AND SZABO}, $j_\mathrm{d'}$ and the simulated flux from the spheres, $j_\mathrm{s}$, multiplied by the number of spheres, $N$:
\begin{equation}
j = j_{d'} + Nj_s
\end{equation}

\begin{figure}[ht]
\begin{center}
\includegraphics[width=\textwidth]{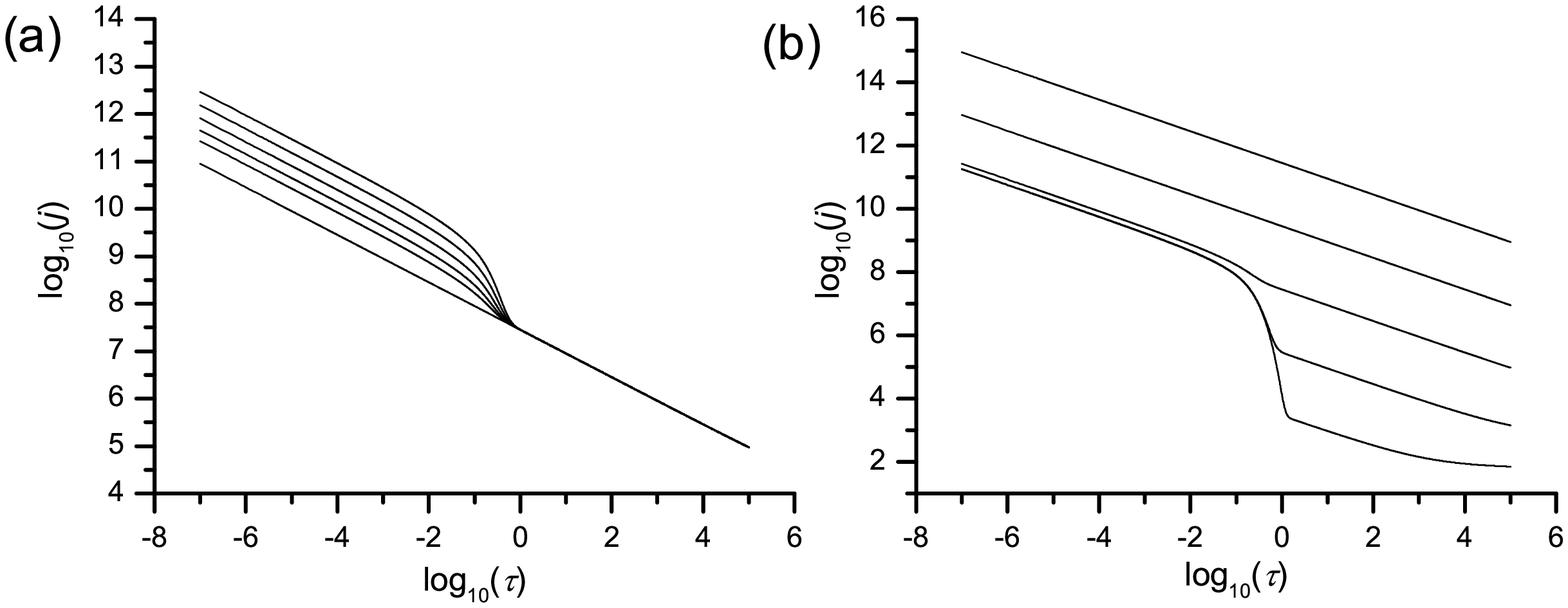}
\caption{(a) Dimensionless chronoamperometric responses of a macrodisc of radius $R_\mathrm{d}=10^4$ with number of hollow spheres $N$ from bottom to top of 0, 1, 2, 4, 8 and 16 $\times 10^{8}$ (b) Dimensionless chronoamperometric responses for $10^8$ spheres of a macrodisc of various radius $R_\mathrm{d}$ of, from bottom to top, $10^2$, $10^3$, $10^4$, $10^5$ and $10^6$.}\label{CHRONO VARY N AND RD}
\end{center}
\end{figure}

Fig \ref{CHRONO VARY N AND RD} (a) shows the calculated chronoamperometric responses for various numbers of hollow spheres on a disc electrode of radius $R_d = 10^4$. The bottom most line represents Equation \ref{SHOUP AND SZABO}, and gives the response for the case with no spheres on the surface. The other lines, from bottom to top, have 1, 2, 4, 8 and 16 $\times 10^8$ spheres respectively. It is seen that, the more spheres there are, the greater the deviation from the Shoup and Szabo case at small $\tau$ values. In all cases however, by the time $\tau$ has reached 1, the contribution from the spheres is insignificant, since all the material inside them has been used up, and only the ``disc'' response remains. 

The response inside the spheres can be expected to have a more significant contribution to the overall current if the experiment is carried out in an ionic liquid (rather than in conventional solvents) due to the drastically lowered diffusion coefficients. For a typical sphere radius of 250 nm\cite{Chen2008, Zhou2010, Chen2012}, and diffusion coefficient in an ionic liquid of $10^{-11}$ m$^2$ s$^{-1}$, this means the contribution from the spheres will not be seen after roughly 5 milliseconds, making it difficult to observe experimentally as it will be masked by double layer charging. Conversely, for aqueous systems, a typical diffusion coefficient would be of the order $10^{-9}$ m$^2$ s$^{-1}$, meaning the signal from inside the spheres is lost after roughly 10 $\mu$s, and will certainly not be observed using conventional chronoamperometry.

Fig \ref{CHRONO VARY N AND RD} (b) shows calculated chronoamperometric responses this time for a fixed number of spheres ($10^8$), but varying $R_d$ from $10^{2}$ (bottom most line), $10^{3}$, $10^{4}$,  $10^{5}$ to $10^{6}$, (top most line). For the smallest disc, the overall response is initially dominated by the spheres. Around $\tau=1$ the response from the spheres dies away as all the material in them is used up. This just leaves the disc response, which eventually reaches a steady state value (due to the small size of the disc). For the largest disc, it is seen that the log-log plot of the response is linear at all times. The current from inside the spheres is swamped by the large disc response and is not seen. The large size of the disc also means that steady state is not reached in this timescale, and a Cottrellian response is all that is seen.

\subsection{Cyclic voltammetry}

\begin{figure}[ht]
\begin{center}
\includegraphics[width=\textwidth]{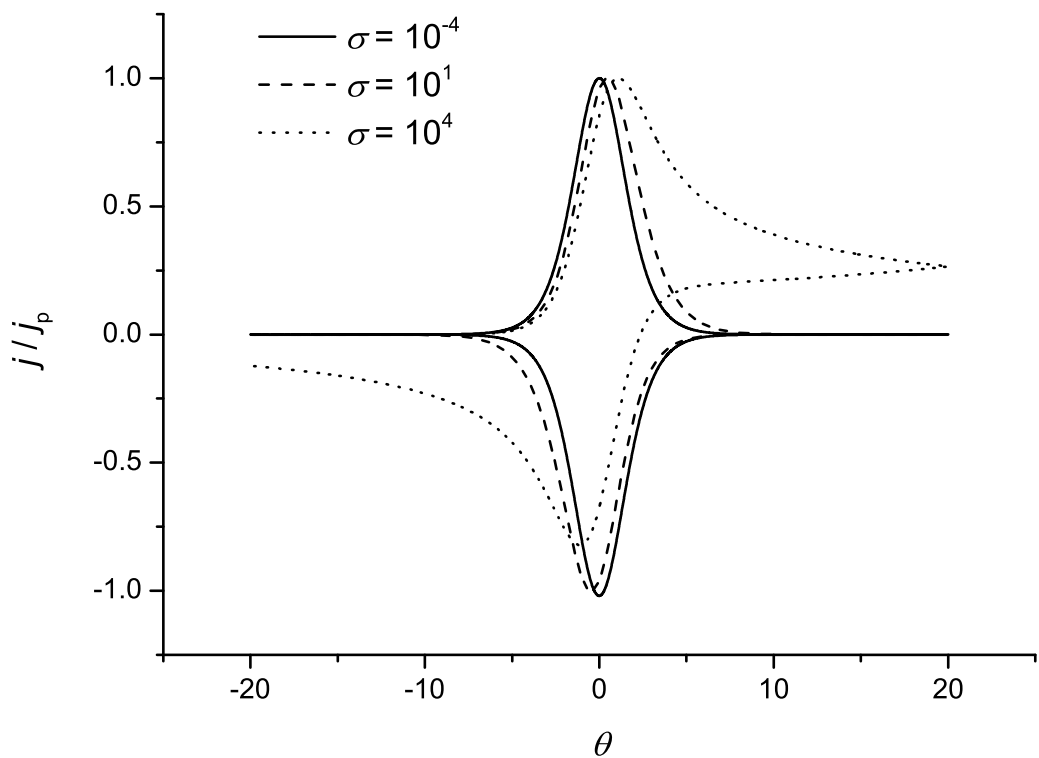}
\caption{Normalised simulated cyclic voltammetric responses from inside hollow spheres at dimensionless scan rates of $10^{-4}$ (solid line), $10^{1}$ (dashed line) and $10^4$ (dotted line). $K^0$ is fixed at $10^{5}$.} \label{DIFFERNET SIGMA}
\end{center}
\end{figure}

Cyclic voltammetry measured inside a sphere will vary greatly in character depending on the scan rate used. At low scan rates (or small spheres) all of the electroactive species inside the sphere will be consumed, leading to a ``thin layer'' response. At high scan rates (or large spheres) the depletion layer cannot extend very far into the interior of the sphere, leading to a ``diffusional'' response. This is seen qualitatively in Fig \ref{DIFFERNET SIGMA} which shows simulated cyclic voltammograms with $\sigma$ values of $10^{-4}$, $10^{1}$ and $10^{4}$ (normalised to a peak height of one). A clear transition is seen from a thin layer response, through an intermediate case to a diffusional response.

\begin{figure}[ht]
\begin{center}
\includegraphics[width=\textwidth]{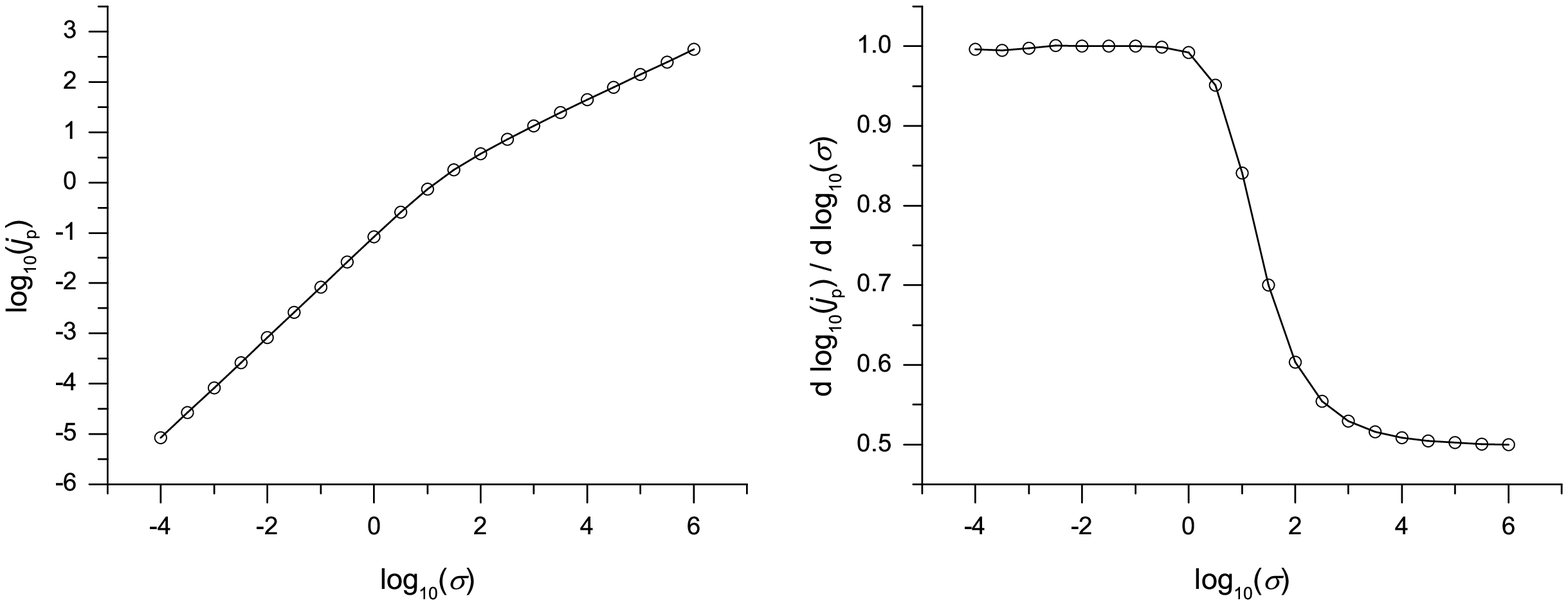}
\caption{(a) Decadic log of dimesnionless peak current simulated inside a sphere as a function of dedadic log of dimensionless scan rate. $K^0$ is set at $10^5$ to ensure complete electrochemical reversibility over the whole range. (b) shows the gradient of (a).} \label{J VS SIGMA}
\end{center}
\end{figure}

This behavior is confirmed with a plot of log$_{10}$($j_p$) (peak dimensionless flux density) against log$_{10}$($\sigma$). For a surface bound/thin layer response, the peak hight is directly proportional to the scan rate. For a diffusional response, the peak hight is proportional to the square root of the scan rate. This is seen in Fig \ref{J VS SIGMA}. Part (a) of this figure is a log-log plot of dimensionless peak height vs $\sigma$, and shows a very clear change of gradient. Part (b) shows the value of this gradient, which changes from one at low $\sigma$ to a half at high $\sigma$.

\begin{figure}[ht]
\begin{center}
\includegraphics[width=\textwidth]{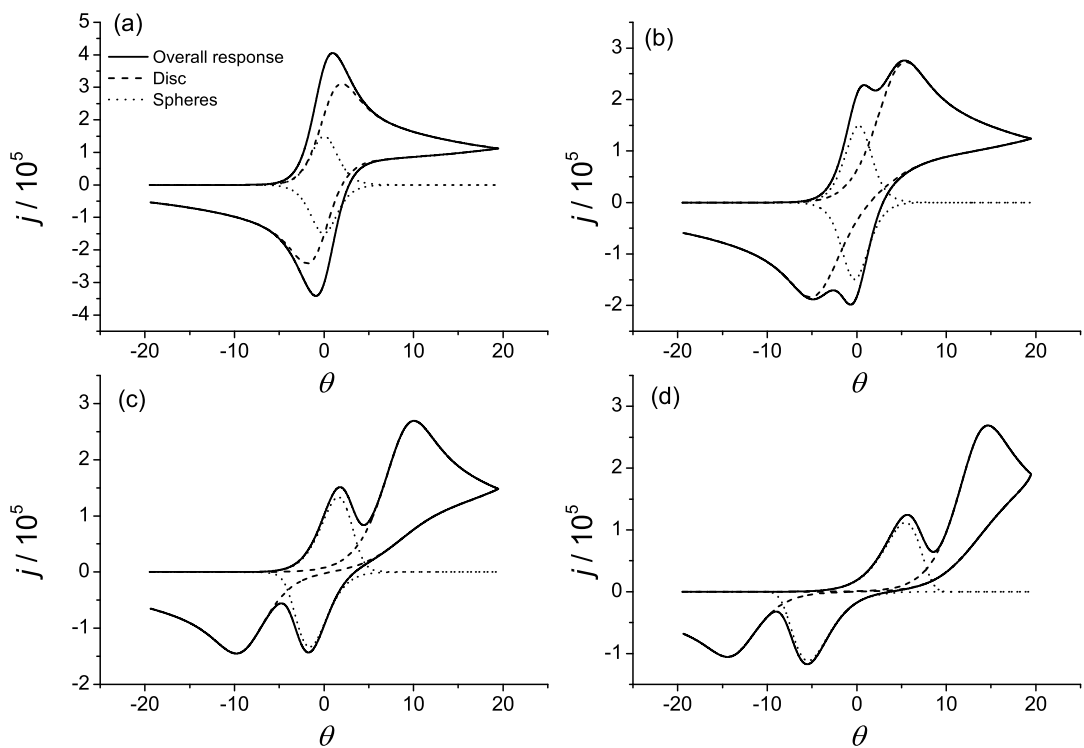}
\caption{Simulated voltammograms at a porous electrode with $\sigma=0.01$, $R_d=4000$, $\alpha=0.5$, $N=1\times10^8$ and heterogeneous rate constants of (a): $1\times10^{-1}$, (b): $1\times10^{-2}$, (c): $1\times10^{-3}$ and (d): $1\times10^{-4}$.} \label{VOLTAMMOGRAMS}
\end{center}
\end{figure}

Simulated cyclic voltammograms at a porous electrode are shown in Fig \ref{VOLTAMMOGRAMS} for dimensionless electrochemical rate constants of (a): $1\times10^{-1}$, (b): $1\times10^{-2}$, (c): $1\times10^{-3}$ and (d): $1\times10^{-4}$. Other parameters are $\sigma=0.01$, $R_d=4000$, $\alpha=0.5$ and $N=1\times 10^8$. It is seen that for the largest electrochemical rate constant, only one peak is observed in each direction of the scan. As the heterogeneous rate constant is lowered however, two peaks emerge, initially close together but getting further apart at smaller $K^0$ values. Also shown in this figure are the relative contributions from electrolysis inside the spheres and from the overall disc (dotted and dashed lines respectively). It is seen that the cause of the peak splitting is the different responses from these two parts of the electrode (hollow voids and exposed surface) to a change in $K^0$. The reasons for this peak splitting are discussed in detail below.

\begin{figure}[ht]
\begin{center}
\includegraphics[width=\textwidth]{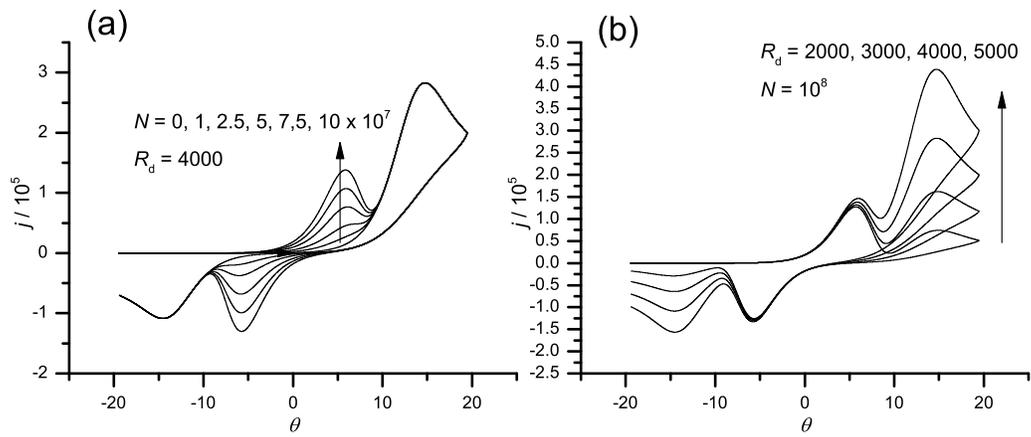}
\caption{Simulated dimensionless voltammograms showing various numbers of hollow spheres in the porous layer (a) and changing the disc electrode radius (b). Other parameters are $\sigma=0.01$, $\alpha=0.5$, $K^0=1\times10^{-4}$} \label{CHANGE N}
\end{center}
\end{figure}

Further to these simulated voltammograms showing the effect of changing $K^0$, Fig. \ref{CHANGE N} shows the effect of changing the number of hollow spheres in the porous electrode (a) and changing the radius of the overall disc electrode (b). Part (a) of this figure shows simulated voltammograms for a porous electrode for parameters in Fig. \ref{VOLTAMMOGRAMS} (d), but with varying numbers of hollow spheres $N$ = 0, 1, 2.5, 5, 7.5 and 10 $\times 10^7$, as indicated on the figure. It is seen that the first wave in the forward scan changes size as the number of spheres is changed, and is absent if no hollow spheres are present. It is also seen that for small numbers of spheres (1.0 and to a lesser extent 2.5 $\times10^7$) then the small pre-peak is difficult to resolve and the precise position of the peak is essentially unreadable. Part (b) of this figure is as part (a), but with $N$ fixed at 10 $^8$ and $R_d$ at 1000, 2000, 3000 and 4000. It is seen that the response due to the overall disc increases as is expected, but for these parameters does not swamp the thin layer signal from inside the spheres. It should be noted however that if $R_d$ becomes extremely large, the thin layer signal will become unresolvable. We below consider the analytical implications of this peak splitting. 

\subsection{The possibility of two peaks arising from a single $\mathrm{A} \pm \mathrm{e^{-}} \rightleftharpoons \mathrm{B}$ process}

Of particular interest is comparing the effect on the voltammetry of changing the heterogeneous rate constant, $K^0$, on the response inside spheres and at a planar macrodisc electrode. Henstridge \emph{et. al.} predicted that at porous film electrodes a single electron transfer process can result in the observation of two peaks in cyclic voltammetry, one from electrolysis within the porous film and the other from semi-infinite linear diffusion at the film surface\cite{Henstridge2012b, Xia2014}. The thin layer response from inside the spheres makes the transition from being electrochemically reversible to irreversible in character at lower heterogeneous rate constants than the diffusional disc response. The fact that this transition from fully reversible to fully irreversible takes place over a different range of heterogeneous rate constants in the two cases means that, for heterogeneous rate constants outside the fully reversible limit, two peaks are seen, one from thin layer electrolysis inside the pores, and one from linear diffusion to the exposed surface. The separation of these peaks is a function of $K^0$, with the peaks becoming more separated at smaller $K^0$, until a limiting separation is reached. This peak splitting may provide a means of directly extracting an \emph{approximate} value for the heterogeneous rate constant and formal potential from experimental data, as described below.

\subsection{Extraction of $K^0$}

\begin{figure}[ht]
\begin{center}
\includegraphics[width=\textwidth]{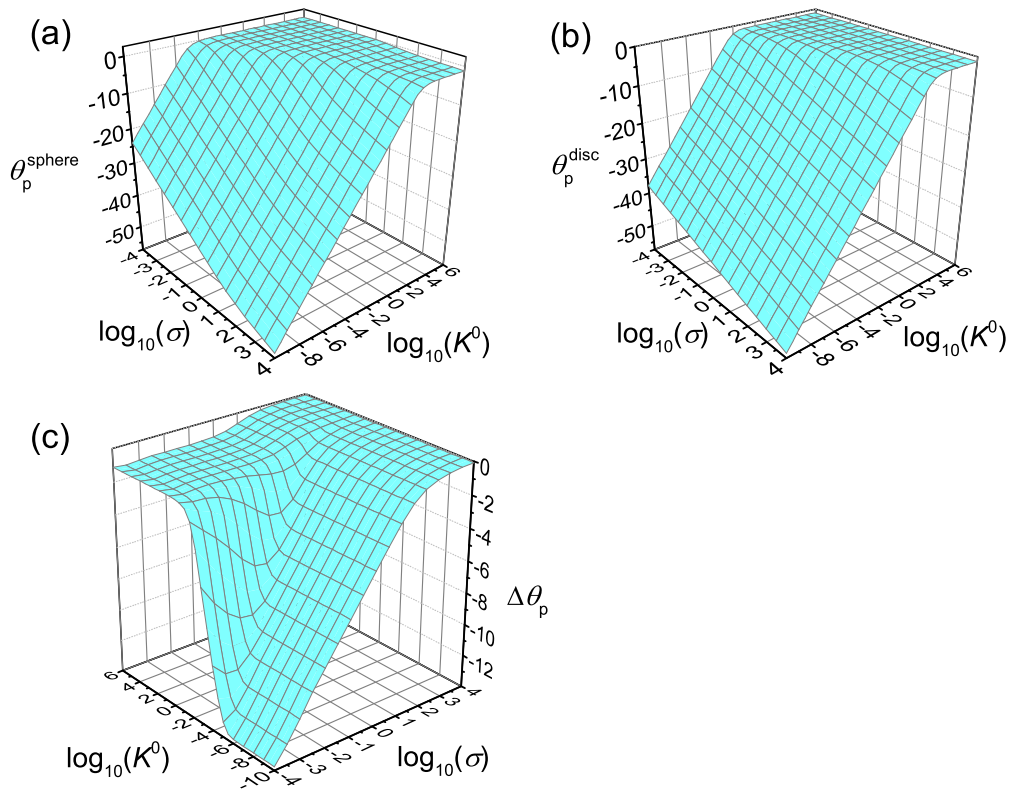}
\caption{Dimensionless peak potential for a range of dimensionless scan rates and heterogeneous rate constants for (a) inside spheres and (b) at macrodiscs. (c) The difference between the two.} \label{THETA P}
\end{center}
\end{figure}

This voltammetric splitting of one peak into two peaks as the electrochemical rate constant is lowered provides an approximate means of estimating $K^0$, which is impossible for a fully irreversible system from a single peak unless the formal potential, $E^\minuso_f$, is known. It is therefore only possible to extract a combined parameter $k^0\mathrm{exp}\left(\frac{\alpha FE^\minuso_f}{RT}\right)$ if only one irreversible peak is seen. If the forward wave contains two peaks however, as here predicted for a porous electrode, then the separation of these two peaks as a function of scan rate may be used to extract an approximate value of $K^0$, and hence $\theta^\minuso_f$, if $\alpha$ is known (or measured). Fig. \ref{THETA P} (a) shows the simulated peak potential on the forward scan for inside spheres ($\theta_\mathrm{p}^\mathrm{sphere}$) plotted as a function of log$_{10}\left(\sigma\right)$ and log$_{10}\left(K^0\right)$. (b) shows the same for a macrodisc ($\theta_\mathrm{p}^\mathrm{disc}$), and (c) shows the difference between the two cases ($\Delta\theta_\mathrm{p}$). In Fig. \ref{THETA P} (c), we see that the separation between the peak responses due to the interior of the spheres and the overall disc is a function of both $\sigma$ and $K^0$. 

For very large values of $\sigma$, the diffusion zone is close to the electrode surface for both inside the spheres and for the disc, thus a diffusional signal is observed for both. In this extreme, there is no peak separation, the electrochemistry is blind to the overall geometry of the electrode at which it occurs.

If $\sigma$ is small but $K^0$ is large, then the response from inside the spheres will be thin layer in nature, and the disc response diffusional, but both responses are fully reversible. This leads to a constant, small peak separation. This corresponds to the simulated voltammograms shown in Fig. \ref{VOLTAMMOGRAMS} (a), the peaks are not separated enough to be resolved into two separate peaks.

If $K^0$ is small, then the peak separation becomes a function of $\sigma$. As $\sigma$ is lowered from very large values, the peak separation moves away from zero (see above) and the peaks begin to shift away from each other more and more, until the separation passes through a maximum and the peaks begin to move closer together again. The point at which this maximum is reached is a function of $K^0$, occurring at lower values of $\sigma$ as $K^0$ becomes smaller.

Noting the above, $K^0$ can in principle be extracted from experimental data obtained over a wide range of scan rates which contain the point of maximum separation described above. The peak separations can be plotted as a function of scan rate and compared to this working surface to estimate $K^0$. $\theta_f^\minuso$ can then also be extracted from the position of the peaks. It should be noted that the data in Fig. \ref{THETA P} is simulated using an $\alpha$ value of 0.5. For experimental data, an $\alpha$ value must be obtained via Tafel analysis and a working surface specific to the obtained $\alpha$ simulated. It should also be noted that this method assumes the ideal geometry of diffusionally independent perfectly spherical pores. This will therefore give us at least a highly approximate estimate of $K^0$ in any real system, where these assumptions are unlikely to be adhered to. In particular the idealised porous electrode structure here simulated is unlikely to be rigorously accurate.

As alluded to above, it should be noted that for small separations in peak potential between the response inside spheres and the response of the disc, the overall response may still contain a single unresolved peak, as in Fig \ref{VOLTAMMOGRAMS} (a), where the responses overlap. The number of spheres present on the surface also has an impact, too few and the thin layer signal will be swamped by the diffusional signal and not visible. This is demonstrated in Fig. \ref{CHANGE N}. It is seen that for a small number of spheres ($N=1\times10^7$ and to a lesser extent $2.5\times10^7$) then the peak position of the first, thin layer signal is not readily readable. Thus for the proposed approximate method of extracting $k^0$, it is desirable to have very clearly defined peaks to read off the peak positions, so an optimal number of spheres on the surface is desirable so two peaks are visible although not so large that the signal from inside the spheres begins to swamp the signal from the disc.

\section{Conclusions}

We have developed a theory which allows insight into the electrochemistry of porous electrode structures formed via electroplating around spherical templates which are then removed. It is concluded that in chronoamperometry the timescale for diffusional transport within the spherical voids of the porous layer is so fast that its contribution is usually masked by double layer charging effects. In cyclic voltammetry however, significant contributions from electrolysis of material both within the voids and from semi-infinite linear diffusion to the electrode surface is predicted to be observed. This mixed contribution of thin layer and semi infinite diffusion is shown, in principle, to be useful in extracting heterogeneous rate constants for quasi reversible and irreversible electron transfer processes.

\section*{Acknowledgments}

For funding, EOB thanks the Engineering \& Physical Sciences Research Council (EPSRC) and St. John's College, Oxford, XJ thanks National Natural Science Foundation of China (20905035) and Jiangsu Overseas Research \& Training Program for University Prominent Young \& Middle-aged Teachers and Presidents, and PL thanks the China Scholarship Council (CSC).

\clearpage

\section*{Figures}

\clearpage

\clearpage

\clearpage

\clearpage

\clearpage

\clearpage

\clearpage

\clearpage

\clearpage

\clearpage

\section*{Tables}

\clearpage

\begin{table}[H]
\begin{center}
\begin{tabular}{l l l l}
\hline
Porous material & Hollow void radius & Target species & Reference\\
\hline
Gold & 320 nm & NADH & \cite{Szamocki2006} \\
\\
Gold & 550 nm& NADH & \cite{Szamocki2007} \\
\\
Ruthenium oxide & 150 nm & NADH & \cite{Lenz2011}\\
\\
Gold & 125 nm & Glucose & \cite{Chen2009}\\
\\
Polyaniline/Prussian Blue & 250 nm & Glucose & \cite{Chen2012}\\
\\
Ionic liquid doped poly(\emph{N}[3-& & &\\
(trimethyoxysilyl)propyl]aniline) & 250 nm & Glucose & \cite{Chen2012a}\\
\\
Platinum nanoparticle & & &\\
modified carbon & 55 nm & Nitrobenzene & \cite{Zhang2012}\\
\\
Gold nanoparticle doped\\
titanium dioxide with\\
immobilised haemoglobin & 250 nm & Hydrogen peroxide & \cite{Wei2011}\\
\\
Gold & 250 nm& C-reactive protein$^\dag$ & \cite{Chen2008} \\
\\
Ionic liquid doped & & &\\
polyanailine & 125 nm & Hepatitis B &\\
& & surface antigen & \cite{Li2009b}\\
\\
Gold & 125 nm& Human Apolipoprotein &\\
& & B-100$^\ddag$ & \cite{Chen2010} \\
\hline
\end{tabular}
\end{center}
\caption{Composition and uses of porous electrodes. $^\dag$ A human protein which is an indicator for inflammation\cite{Whicher1999}. $^\ddag$ A human protein which is an indicator for coronary artery disease\cite{Haidari2001}.}
\label{POROUS USES}
\end{table}

\clearpage

\begin{table}
\begin{center}
\begin{tabular}{l l l}
\hline
Parameter & Description & Units\\
\hline
$\alpha$/$\beta$ & Electron transfer coefficients & Unitless\\
\\
$c_\mathrm{i}$ & Concentration of species i & mol m$^{-3}$\\
\\
$c_\mathrm{i}^{*}$ & Bulk solution concentration of species i & mol m$^{-3}$\\
\\
$c_\mathrm{i}^{0}$ & Electrode surface concentration of species i & mol m$^{-3}$\\
\\
$D_\mathrm{i}$ & Diffusion coefficient of species i & m$^{2}$ s$^{-1}$\\
\\
$E$ & Applied potential & V\\
\\
$E^0_\mathrm{f}$ & Formal potential & V\\
\\
$I$ & Current & A\\
\\
$k^0$ & Heterogeneous rate constant & m s$^{-1}$\\
\\
$r$ & radial coordinate & m\\
\\
$r_{s/d}$ & Radius of sphere/disc & m\\
\\
$N$ & Number of spheres & Unitless\\
\\
$t$ & Time & s\\
\\
$z$ & $z$ coordinate & m\\
\hline
\end{tabular}
\end{center}
\caption{Parameter definitions}
\label{DIMENSIONAL}
\end{table}

\clearpage

\begin{table}
\begin{center}
\begin{tabular}{l l}
\hline
Normalised Parameter & Definition\\
\hline
$C_\mathrm{i}$ & $\frac{c_\mathrm{i}}{c^*_\mathrm{A}}$\\
\\
$D^{'}_\mathrm{i}$ & $\frac{D_\mathrm{i}}{D_\mathrm{A}}$\\
\\
$K^0$ & $\frac{r_\mathrm{s}}{D_\mathrm{A}}k^0$\\
\\
$R$ & $\frac{R}{r_\mathrm{s}}$\\
\\
$\sigma$ & $\frac{Fr_\mathrm{s}^2}{RTD_\mathrm{A}}\nu$\\
\\
$\theta$ & $\frac{F\left(E-E^0_\mathrm{f}\right)}{RT}$\\
\\
$\tau$ & $\frac{D_\mathrm{A}}{r_s^2}t$\\
\\
$Z$ & $\frac{z}{r_\mathrm{d}}$\\
\\
\hline
\end{tabular}
\end{center}
\caption{Normalised parameter definitions}
\label{DIMENSIONLESS}
\end{table}

\clearpage

\begin{table}
\begin{center}
\begin{tabular}{c c c}
\hline
Boundary & Chronoamperometry condition & Cyclic voltammetry condition\\
\hline
All $R$, $\tau < 0$ & {$C_\mathrm{A} = 1$} & {$C_\mathrm{A} = 1$}\\
\\
 & {$C_\mathrm{B} = 0$} & {$C_\mathrm{B} = 0$}\\
\\
\\
$R=1$, $\tau \geq 0$ & $C_\mathrm{A} = 0$ & $\frac{\partial{C_\mathrm{A}}}{\partial{R}} = K^0\left[C_\mathrm{A}^0\text{e}^{-\alpha\theta} - C_\mathrm{B}^0\text{e}^{\beta\theta}\right]$\\
\\
 & {$D^{'}_\mathrm{B}\frac{\partial{C_\mathrm{B}}}{\partial{R}} = -\frac{\partial{C_\mathrm{A}}}{\partial{R}}$} & {$D^{'}_\mathrm{B}\frac{\partial{C_\mathrm{B}}}{\partial{R}} = -\frac{\partial{C_\mathrm{A}}}{\partial{R}}$}\\
\\
\\
$R=0$, $\tau \geq 0$ & {$\frac{\partial{C_\mathrm{i}}}{\partial{R}} = 0$} & {$\frac{\partial{C_\mathrm{i}}}{\partial{R}} = 0$}\\
\\
\hline
\end{tabular}
\end{center}
\caption{Normalised boundary conditions for electrochemistry inside a hollow sphere}
\label{SPHERE BOUNDARY CONDITIONS}
\end{table}

\clearpage

\begin{table}
\begin{center}
\begin{tabular}{c c}
\hline
Boundary & Cyclic voltammetry condition\\
\hline
All $Z$, $\tau < 0$ & $C_\mathrm{A} = 1$\\
\\
 & $C_\mathrm{B} = 0$\\
\\
\\
$R=0$, $\tau \geq 0$ & $\frac{\partial{C_\mathrm{A}}}{\partial{R}} = K^0\left[C_\mathrm{A}^0\text{e}^{-\alpha\theta} - C_\mathrm{B}^0\text{e}^{\beta\theta}\right]$\\
\\
 & $D^{'}_\mathrm{B}\frac{\partial{C_\mathrm{B}}}{\partial{R}} = -\frac{\partial{C_\mathrm{A}}}{\partial{R}}$\\
\\
\\
$R=6\sqrt{D^{'}_\mathrm{max}\tau_\mathrm{max}}$, $\tau \geq 0$ & $\frac{\partial{C_\mathrm{i}}}{\partial{R}} = 0$\\
\\
\hline
\end{tabular}
\end{center}
\caption{Normalised boundary conditions for cyclic voltammetry at a macrodisc electrode}
\label{DISC BOUNDARY CONDITIONS}
\end{table}

\clearpage

\providecommand*\mcitethebibliography{\thebibliography}
\csname @ifundefined\endcsname{endmcitethebibliography}
  {\let\endmcitethebibliography\endthebibliography}{}


\begin{mcitethebibliography}{40}
\providecommand*\natexlab[1]{#1}
\providecommand*\mciteSetBstSublistMode[1]{}
\providecommand*\mciteSetBstMaxWidthForm[2]{}
\providecommand*\mciteBstWouldAddEndPuncttrue
  {\def\EndOfBibitem{\unskip.}}
\providecommand*\mciteBstWouldAddEndPunctfalse
  {\let\EndOfBibitem\relax}
\providecommand*\mciteSetBstMidEndSepPunct[3]{}
\providecommand*\mciteSetBstSublistLabelBeginEnd[3]{}
\providecommand*\EndOfBibitem{}
\mciteSetBstSublistMode{f}
\mciteSetBstMaxWidthForm{subitem}{(\alph{mcitesubitemcount})}
\mciteSetBstSublistLabelBeginEnd
  {\mcitemaxwidthsubitemform\space}
  {\relax}
  {\relax}

\bibitem[Szamocki et~al.(2006)Szamocki, Reculusa, Ravaine, Bartlett, Kuhn, and
  Hempelmann]{Szamocki2006}
Szamocki,~R.; Reculusa,~S.; Ravaine,~S.; Bartlett,~P.~N.; Kuhn,~A.;
  Hempelmann,~R. \emph{Angew. Chem. Int. Ed.} \textbf{2006}, \emph{45}, 1317 --
  1321\relax
\mciteBstWouldAddEndPuncttrue
\mciteSetBstMidEndSepPunct{\mcitedefaultmidpunct}
{\mcitedefaultendpunct}{\mcitedefaultseppunct}\relax
\EndOfBibitem
\bibitem[Szamocki et~al.(2007)Szamocki, Velichko, Holzapfel, M\"{u}cklich,
  Ravaine, Garrigue, Sojic, Hempelmann, and Kuhn]{Szamocki2007}
Szamocki,~R.; Velichko,~A.; Holzapfel,~C.; M\"{u}cklich,~F.; Ravaine,~S.;
  Garrigue,~P.; Sojic,~N.; Hempelmann,~R.; Kuhn,~A. \emph{Anal. Chem.}
  \textbf{2007}, \emph{79}, 533 -- 539\relax
\mciteBstWouldAddEndPuncttrue
\mciteSetBstMidEndSepPunct{\mcitedefaultmidpunct}
{\mcitedefaultendpunct}{\mcitedefaultseppunct}\relax
\EndOfBibitem
\bibitem[Lenz et~al.(2011)Lenz, Trieu, Hempelmann, and Kuhn]{Lenz2011}
Lenz,~J.; Trieu,~V.; Hempelmann,~R.; Kuhn,~A. \emph{Electroanalysis}
  \textbf{2011}, \emph{23}, 1186 -- 1192\relax
\mciteBstWouldAddEndPuncttrue
\mciteSetBstMidEndSepPunct{\mcitedefaultmidpunct}
{\mcitedefaultendpunct}{\mcitedefaultseppunct}\relax
\EndOfBibitem
\bibitem[Chen et~al.(2009)Chen, Xuan, Jiang, and Zhu]{Chen2009}
Chen,~X.; Xuan,~J.; Jiang,~L.; Zhu,~J. \emph{Sci. in China B} \textbf{2009},
  \emph{52}, 1999 -- 2005\relax
\mciteBstWouldAddEndPuncttrue
\mciteSetBstMidEndSepPunct{\mcitedefaultmidpunct}
{\mcitedefaultendpunct}{\mcitedefaultseppunct}\relax
\EndOfBibitem
\bibitem[Chen et~al.(2012)Chen, Chen, Tian, Yan, and Yao]{Chen2012}
Chen,~X.; Chen,~Z.; Tian,~R.; Yan,~W.; Yao,~C. \emph{Anal. Chim. Acta}
  \textbf{2012}, \emph{723}, 94 -- 100\relax
\mciteBstWouldAddEndPuncttrue
\mciteSetBstMidEndSepPunct{\mcitedefaultmidpunct}
{\mcitedefaultendpunct}{\mcitedefaultseppunct}\relax
\EndOfBibitem
\bibitem[Chen et~al.(2012)Chen, Zhu, Tian, and Yao]{Chen2012a}
Chen,~X.; Zhu,~J.; Tian,~R.; Yao,~C. \emph{Sens. Actuators B} \textbf{2012},
  \emph{163}, 272 -- 280\relax
\mciteBstWouldAddEndPuncttrue
\mciteSetBstMidEndSepPunct{\mcitedefaultmidpunct}
{\mcitedefaultendpunct}{\mcitedefaultseppunct}\relax
\EndOfBibitem
\bibitem[Zhang et~al.(2012)Zhang, Zeng, Bo, Wang, and Guo]{Zhang2012}
Zhang,~Y.; Zeng,~L.; Bo,~X.; Wang,~H.; Guo,~L. \emph{Anal. Chim. Acta}
  \textbf{2012}, \emph{752}, 45 -- 52\relax
\mciteBstWouldAddEndPuncttrue
\mciteSetBstMidEndSepPunct{\mcitedefaultmidpunct}
{\mcitedefaultendpunct}{\mcitedefaultseppunct}\relax
\EndOfBibitem
\bibitem[Wei et~al.(2011)Wei, Xin, Du, and Li]{Wei2011}
Wei,~N.; Xin,~X.; Du,~J.; Li,~J. \emph{Biosens. Biolelectron.} \textbf{2011},
  \emph{26}, 3602 -- 3607\relax
\mciteBstWouldAddEndPuncttrue
\mciteSetBstMidEndSepPunct{\mcitedefaultmidpunct}
{\mcitedefaultendpunct}{\mcitedefaultseppunct}\relax
\EndOfBibitem
\bibitem[Chen et~al.(2008)Chen, Wang, Zhou, Yan, Li, and Zhu]{Chen2008}
Chen,~X.; Wang,~Y.; Zhou,~J.; Yan,~W.; Li,~X.; Zhu,~J. \emph{Anal. Chem.}
  \textbf{2008}, \emph{80}, 2133 -- 2140\relax
\mciteBstWouldAddEndPuncttrue
\mciteSetBstMidEndSepPunct{\mcitedefaultmidpunct}
{\mcitedefaultendpunct}{\mcitedefaultseppunct}\relax
\EndOfBibitem
\bibitem[Li et~al.(2009)Li, Dai, Liu, Chen, Yan, Chen, and Zhu]{Li2009b}
Li,~X.; Dai,~L.; Liu,~Y.; Chen,~X.; Yan,~W.; Chen,~X.; Zhu,~J. \emph{Adv.
  Funct. Mater.} \textbf{2009}, \emph{19}, 3120 -- 3128\relax
\mciteBstWouldAddEndPuncttrue
\mciteSetBstMidEndSepPunct{\mcitedefaultmidpunct}
{\mcitedefaultendpunct}{\mcitedefaultseppunct}\relax
\EndOfBibitem
\bibitem[Chen et~al.(2010)Chen, Zhou, Xuan, Yan, Jiang, and Zhu]{Chen2010}
Chen,~X.; Zhou,~J.; Xuan,~J.; Yan,~W.; Jiang,~L.; Zhu,~J. \emph{Analyst}
  \textbf{2010}, \emph{135}, 2629--2636\relax
\mciteBstWouldAddEndPuncttrue
\mciteSetBstMidEndSepPunct{\mcitedefaultmidpunct}
{\mcitedefaultendpunct}{\mcitedefaultseppunct}\relax
\EndOfBibitem
\bibitem[Streeter et~al.(2008)Streeter, Wildgoose, Shao, and
  Compton]{Streeter2008c}
Streeter,~I.; Wildgoose,~G.~G.; Shao,~L.; Compton,~R.~G. \emph{Sens. Actuators,
  B} \textbf{2008}, \emph{133}, 462 -- 466\relax
\mciteBstWouldAddEndPuncttrue
\mciteSetBstMidEndSepPunct{\mcitedefaultmidpunct}
{\mcitedefaultendpunct}{\mcitedefaultseppunct}\relax
\EndOfBibitem
\bibitem[Sims et~al.(2010)Sims, Rees, Dickinson, and Compton]{Sims2010}
Sims,~M.~J.; Rees,~N.~V.; Dickinson,~E. J.~F.; Compton,~R.~G. \emph{Sens.
  Actuators B} \textbf{2010}, \emph{144}, 153 -- 158\relax
\mciteBstWouldAddEndPuncttrue
\mciteSetBstMidEndSepPunct{\mcitedefaultmidpunct}
{\mcitedefaultendpunct}{\mcitedefaultseppunct}\relax
\EndOfBibitem
\bibitem[Henstridge et~al.(2010)Henstridge, Dickinson, Aslanoglu,
  Batchelor-McAuley, and Compton]{Henstridge2010}
Henstridge,~M.~C.; Dickinson,~E. J.~F.; Aslanoglu,~M.; Batchelor-McAuley,~C.;
  Compton,~R.~G. \emph{Sens. Actuators B} \textbf{2010}, \emph{145}, 417 --
  427\relax
\mciteBstWouldAddEndPuncttrue
\mciteSetBstMidEndSepPunct{\mcitedefaultmidpunct}
{\mcitedefaultendpunct}{\mcitedefaultseppunct}\relax
\EndOfBibitem
\bibitem[Punckt et~al.(2013)Punckt, Pope, and Aksay]{Punckt2013}
Punckt,~C.; Pope,~M.~A.; Aksay,~I.~A. \emph{J. Phys. Chem. C} \textbf{2013},
  \emph{117}, 16076 -- 16086\relax
\mciteBstWouldAddEndPuncttrue
\mciteSetBstMidEndSepPunct{\mcitedefaultmidpunct}
{\mcitedefaultendpunct}{\mcitedefaultseppunct}\relax
\EndOfBibitem
\bibitem[Brookes et~al.(2003)Brookes, Davies, Fisher, Evans, Wilkins, Yunus,
  Wadhawan, and Compton]{Brookes2003}
Brookes,~B.~A.; Davies,~T.~J.; Fisher,~A.~C.; Evans,~R.~G.; Wilkins,~S.~J.;
  Yunus,~K.; Wadhawan,~J.~D.; Compton,~R.~G. \emph{J. Phys. Chem. B}
  \textbf{2003}, \emph{107}, 1616 -- 1627\relax
\mciteBstWouldAddEndPuncttrue
\mciteSetBstMidEndSepPunct{\mcitedefaultmidpunct}
{\mcitedefaultendpunct}{\mcitedefaultseppunct}\relax
\EndOfBibitem
\bibitem[Davies et~al.(2004)Davies, Moore, Banks, and Compton]{Davies2004}
Davies,~T.~J.; Moore,~R.~R.; Banks,~C.~E.; Compton,~R.~G. \emph{J. Electroanal.
  Chem.} \textbf{2004}, \emph{574}, 123 -- 152\relax
\mciteBstWouldAddEndPuncttrue
\mciteSetBstMidEndSepPunct{\mcitedefaultmidpunct}
{\mcitedefaultendpunct}{\mcitedefaultseppunct}\relax
\EndOfBibitem
\bibitem[Davies et~al.(2005)Davies, Banks, and Compton]{Davies2005b}
Davies,~T.~J.; Banks,~C.~E.; Compton,~R.~G. \emph{J. Solid State Electrochem.}
  \textbf{2005}, \emph{9}, 797 -- 808\relax
\mciteBstWouldAddEndPuncttrue
\mciteSetBstMidEndSepPunct{\mcitedefaultmidpunct}
{\mcitedefaultendpunct}{\mcitedefaultseppunct}\relax
\EndOfBibitem
\bibitem[Chevallier et~al.(2005)Chevallier, Davies, Klymenko, Jiang, Jones, and
  Compton]{Chevallier2005}
Chevallier,~F.~G.; Davies,~T.~J.; Klymenko,~O.~V.; Jiang,~L.; Jones,~T. G.~J.;
  Compton,~R.~G. \emph{J. Electroanal. Chem.} \textbf{2005}, \emph{577}, 211 --
  221\relax
\mciteBstWouldAddEndPuncttrue
\mciteSetBstMidEndSepPunct{\mcitedefaultmidpunct}
{\mcitedefaultendpunct}{\mcitedefaultseppunct}\relax
\EndOfBibitem
\bibitem[Davies et~al.(2003)Davies, Brookes, Fisher, Yunus, Wilkins, Greene,
  Wadhawan, and Compton]{Davies2003}
Davies,~T.~J.; Brookes,~B.~A.; Fisher,~A.~C.; Yunus,~K.; Wilkins,~S.~J.;
  Greene,~P.~R.; Wadhawan,~J.~D.; Compton,~R.~G. \emph{J. Phys. Chem. B}
  \textbf{2003}, \emph{107}, 6431 -- 6444\relax
\mciteBstWouldAddEndPuncttrue
\mciteSetBstMidEndSepPunct{\mcitedefaultmidpunct}
{\mcitedefaultendpunct}{\mcitedefaultseppunct}\relax
\EndOfBibitem
\bibitem[Davies and Compton(2005)Davies, and Compton]{Davies2005}
Davies,~T.~J.; Compton,~R.~G. \emph{J. Electroanal. Chem.} \textbf{2005},
  \emph{585}, 63 -- 82\relax
\mciteBstWouldAddEndPuncttrue
\mciteSetBstMidEndSepPunct{\mcitedefaultmidpunct}
{\mcitedefaultendpunct}{\mcitedefaultseppunct}\relax
\EndOfBibitem
\bibitem[Davies et~al.(2005)Davies, Ward-Jones, Banks, Del~Campo, Mas, Munoz,
  and Compton]{Davies2005a}
Davies,~T.~J.; Ward-Jones,~S.; Banks,~C.~E.; Del~Campo,~J.; Mas,~R.;
  Munoz,~F.~X.; Compton,~R.~G. \emph{J. Electroanal. Chem.} \textbf{2005},
  \emph{585}, 51 -- 62\relax
\mciteBstWouldAddEndPuncttrue
\mciteSetBstMidEndSepPunct{\mcitedefaultmidpunct}
{\mcitedefaultendpunct}{\mcitedefaultseppunct}\relax
\EndOfBibitem
\bibitem[Butler(1924)]{Butler1924}
Butler,~J. A.~V. \emph{Trans. Faraday Soc.} \textbf{1924}, \emph{19}, 729 --
  733\relax
\mciteBstWouldAddEndPuncttrue
\mciteSetBstMidEndSepPunct{\mcitedefaultmidpunct}
{\mcitedefaultendpunct}{\mcitedefaultseppunct}\relax
\EndOfBibitem
\bibitem[Laborda et~al.(2013)Laborda, Henstridge, Batchelor-McAuley, and
  Compton]{Laborda2013}
Laborda,~E.; Henstridge,~M.~C.; Batchelor-McAuley,~C.; Compton,~R.~G.
  \emph{Chem. Soc. Rev.} \textbf{2013}, \emph{42}, 4894 -- 4905\relax
\mciteBstWouldAddEndPuncttrue
\mciteSetBstMidEndSepPunct{\mcitedefaultmidpunct}
{\mcitedefaultendpunct}{\mcitedefaultseppunct}\relax
\EndOfBibitem
\bibitem[Shoup and Szabo(1982)Shoup, and Szabo]{Shoup1982}
Shoup,~D.; Szabo,~A. \emph{J. Electroanal. Chem.} \textbf{1982}, \emph{140},
  237 -- 245\relax
\mciteBstWouldAddEndPuncttrue
\mciteSetBstMidEndSepPunct{\mcitedefaultmidpunct}
{\mcitedefaultendpunct}{\mcitedefaultseppunct}\relax
\EndOfBibitem
\bibitem[Paddon et~al.(2007)Paddon, Silvester, Bhatti, Donohoe, and
  Compton]{Paddon2007}
Paddon,~C.~A.; Silvester,~D.~S.; Bhatti,~F.~L.; Donohoe,~T.~J.; Compton,~R.~G.
  \emph{Electroanalysis} \textbf{2007}, \emph{19}, 11 -- 22\relax
\mciteBstWouldAddEndPuncttrue
\mciteSetBstMidEndSepPunct{\mcitedefaultmidpunct}
{\mcitedefaultendpunct}{\mcitedefaultseppunct}\relax
\EndOfBibitem
\bibitem[Klymenko et~al.(2004)Klymenko, Evans, Hardacre, Svir, and
  Compton]{Klymenko2004}
Klymenko,~O.~V.; Evans,~R.~G.; Hardacre,~C.; Svir,~I.~B.; Compton,~R.~G.
  \emph{J. Electroanal. Chem.} \textbf{2004}, \emph{571}, 211 -- 221\relax
\mciteBstWouldAddEndPuncttrue
\mciteSetBstMidEndSepPunct{\mcitedefaultmidpunct}
{\mcitedefaultendpunct}{\mcitedefaultseppunct}\relax
\EndOfBibitem
\bibitem[Gavaghan(1998)]{Gavaghan1998}
Gavaghan,~D.~J. \emph{J. Electroanal. Chem.} \textbf{1998}, \emph{456}, 25 --
  35\relax
\mciteBstWouldAddEndPuncttrue
\mciteSetBstMidEndSepPunct{\mcitedefaultmidpunct}
{\mcitedefaultendpunct}{\mcitedefaultseppunct}\relax
\EndOfBibitem
\bibitem[Gavaghan(1998)]{Gavaghan1998a}
Gavaghan,~D.~J. \emph{J. Electroanal. Chem.} \textbf{1998}, \emph{456}, 13 --
  23\relax
\mciteBstWouldAddEndPuncttrue
\mciteSetBstMidEndSepPunct{\mcitedefaultmidpunct}
{\mcitedefaultendpunct}{\mcitedefaultseppunct}\relax
\EndOfBibitem
\bibitem[Gavaghan(1998)]{Gavaghan1998b}
Gavaghan,~D.~J. \emph{J. Electroanal. Chem.} \textbf{1998}, \emph{456}, 1 --
  12\relax
\mciteBstWouldAddEndPuncttrue
\mciteSetBstMidEndSepPunct{\mcitedefaultmidpunct}
{\mcitedefaultendpunct}{\mcitedefaultseppunct}\relax
\EndOfBibitem
\bibitem[Crank and Nicolson(1947)Crank, and Nicolson]{Crank1947}
Crank,~J.; Nicolson,~E. \emph{Proc. Camb. Phil. Soc.} \textbf{1947}, \emph{43},
  50 -- 67\relax
\mciteBstWouldAddEndPuncttrue
\mciteSetBstMidEndSepPunct{\mcitedefaultmidpunct}
{\mcitedefaultendpunct}{\mcitedefaultseppunct}\relax
\EndOfBibitem
\bibitem[Press et~al.(2007)Press, Teukolsky, Vetterling, and
  Flannery]{Press2007}
Press,~W.~H., Teukolsky,~S.~A., Vetterling,~W.~T., Flannery,~B.~P., Eds.
  \emph{Numerical Recipes: The Art of Scientific Computing}; Cambridge
  University Press, 2007\relax
\mciteBstWouldAddEndPuncttrue
\mciteSetBstMidEndSepPunct{\mcitedefaultmidpunct}
{\mcitedefaultendpunct}{\mcitedefaultseppunct}\relax
\EndOfBibitem
\bibitem[Cottrell(1902)]{Cottrell1902}
Cottrell,~F.~G. \emph{Z. Physik. Chem.} \textbf{1902}, \emph{44}, 385 --
  431\relax
\mciteBstWouldAddEndPuncttrue
\mciteSetBstMidEndSepPunct{\mcitedefaultmidpunct}
{\mcitedefaultendpunct}{\mcitedefaultseppunct}\relax
\EndOfBibitem
\bibitem[Compton and Banks(2010)Compton, and Banks]{Compton2010}
Compton,~R.~G.; Banks,~C.~E. \emph{Understanding Voltammetry}, 2nd ed.; World
  Scientific: Singapore, 2010\relax
\mciteBstWouldAddEndPuncttrue
\mciteSetBstMidEndSepPunct{\mcitedefaultmidpunct}
{\mcitedefaultendpunct}{\mcitedefaultseppunct}\relax
\EndOfBibitem
\bibitem[Zhou et~al.(2010)Zhou, Huang, Xuan, Zhang, and Zhu]{Zhou2010}
Zhou,~J.; Huang,~H.; Xuan,~J.; Zhang,~J.; Zhu,~J. \emph{Biosens. Bioelectron.}
  \textbf{2010}, \emph{26}, 834 -- 840\relax
\mciteBstWouldAddEndPuncttrue
\mciteSetBstMidEndSepPunct{\mcitedefaultmidpunct}
{\mcitedefaultendpunct}{\mcitedefaultseppunct}\relax
\EndOfBibitem
\bibitem[Henstridge et~al.(2012)Henstridge, Dickinson, and
  Compton]{Henstridge2012b}
Henstridge,~M.~C.; Dickinson,~E. J.~F.; Compton,~R.~G. \emph{Russ. J.
  Electrochem.} \textbf{2012}, \emph{48}, 6329 -- 2333\relax
\mciteBstWouldAddEndPuncttrue
\mciteSetBstMidEndSepPunct{\mcitedefaultmidpunct}
{\mcitedefaultendpunct}{\mcitedefaultseppunct}\relax
\EndOfBibitem
\bibitem[Xia et~al.(2014)Xia, Yu, Hu, Feng, Chen, Shi, and Weng]{Xia2014}
Xia,~S.; Yu,~M.; Hu,~J.; Feng,~J.; Chen,~J.; Shi,~M.; Weng,~X.
  \emph{Electrochem. Commun.} \textbf{2014}, \emph{40}, 67 -- 70\relax
\mciteBstWouldAddEndPuncttrue
\mciteSetBstMidEndSepPunct{\mcitedefaultmidpunct}
{\mcitedefaultendpunct}{\mcitedefaultseppunct}\relax
\EndOfBibitem
\bibitem[Whicher et~al.(1999)Whicher, Biasucci, and Rifai]{Whicher1999}
Whicher,~J.; Biasucci,~L.; Rifai,~N. \emph{Clin. Chem. Lab. Med.}
  \textbf{1999}, \emph{37}, 495 -- 503\relax
\mciteBstWouldAddEndPuncttrue
\mciteSetBstMidEndSepPunct{\mcitedefaultmidpunct}
{\mcitedefaultendpunct}{\mcitedefaultseppunct}\relax
\EndOfBibitem
\bibitem[Haidari et~al.(2001)Haidari, Moghadam, Chinicar, Ahmadieh, and
  Doosti]{Haidari2001}
Haidari,~M.; Moghadam,~M.; Chinicar,~M.; Ahmadieh,~A.; Doosti,~M. \emph{Clin.
  Biochem.} \textbf{2001}, \emph{34}, 149 -- 155\relax
\mciteBstWouldAddEndPuncttrue
\mciteSetBstMidEndSepPunct{\mcitedefaultmidpunct}
{\mcitedefaultendpunct}{\mcitedefaultseppunct}\relax
\EndOfBibitem
\end{mcitethebibliography}
\end{document}